\documentclass[%
aps, %
pra, 
preprint, onecolumn, nofootinbib, 11pt, 
citeautoscript, %
a4paper, %
eqsecnum, 
amsfonts, amsmath, amssymb]{revtex4-1}
\usepackage[sort&compress]{natbib}
\usepackage{txfonts,anyfontsize}
\usepackage{ascmac}
\usepackage{mathrsfs}
\usepackage{bm}
\usepackage{color}
\usepackage{graphicx}
\usepackage{bm}
\usepackage{braket}
\usepackage{mathrsfs}
\usepackage[normalem]{ulem}
\usepackage{hhline}
\usepackage{hyperref}
\hypersetup{%
	colorlinks=true, %
	citecolor=blue, %
	pdfstartview={FitH} %
}
\usepackage[capitalise]{cleveref} 
\newcommand{\Ni}{n_{\mathrm{i}}}

\newcommand{\br}{\bm{r}}
\newcommand{\bk}{\bm{k}}
\newcommand{\bp}{\bm{p}}

\newcommand{\zi}{i}
\newcommand{\dd}[1]{\mathrm{d} #1\,}
\renewcommand{\Im}{\mathrm{Im}\,}
\newcommand{\tr}{\mathrm{tr}\,}

\newcommand{\Ord}[1]{\mathcal{O}(#1)}
\newcommand{\R}{\mathrm{R}}
\newcommand{\A}{\mathrm{A}}

\newcommand{\bbracket}[1]{\bigl\langle #1 \bigr\rangle}
\newcommand{\bbracketld}[1]{\bigl\langle #1 \bigr\rangle_{\mathrm{ld}}}
\newcommand{\bbracketsurf}[1]{\bigl\langle #1 \bigr\rangle_{\mathrm{surf}}}
\graphicspath{{./}}
\begin{document}
%
\title{Transport Coefficients of Dirac Ferromagnet: Effects of Vertex Corrections}
\date{\today}
\author{Junji Fujimoto}
\email[E-mail address: ]{fujimoto.junji.8s@kyoto-u.ac.jp}
\affiliation{Institute for Chemical Research, Kyoto University, Uji, Kyoto 611-0011, Japan}
\affiliation{RIKEN Center for Emergent Matter Science, Wako, Saitama 351-0198, Japan}
\begin{abstract}
 As a strongly spin-orbit coupled metallic model with ferromagnetism, we have considered an extended Stoner model to the relativistic regime, named Dirac ferromagnet in three dimensions.
 In the previous paper~[Phys. Rev. B \textbf{90}, 214418 (2014)], we studied the transport properties giving rise to the anisotropic magnetoresistance~(AMR) and the anomalous Hall effect~(AHE) with the impurity potential being taken into account only as the self-energy.
 The effects of the vertex corrections~(VCs) to AMR and AHE are reported in this paper.
 AMR is found not to change quantitatively when the VCs is considered, although the transport lifetime is different from the one-electron lifetime and the charge current includes additional contributions from the correlation with spin currents.
 The side-jump and the skew-scattering contributions to AHE are also calculated.
 The skew-scattering contribution is dominant in the clean case as can be seen in the spin Hall effect in the non-magnetic Dirac electron system.
\end{abstract}
\maketitle

\section{\label{sec:intro}Introduction}
 The spin-orbit coupling~(SOC) yields a variety of phenomena in ferromagnetic materials, such as the anisotropic magnetoresistance~(AMR)~\cite{Thomson1857,Smit1951,Campbell1970,Potter1974,McGuire1975,Rushforth2007,Kato2008,Vyborny2009,Kovalev2009,Wimmer2014}, the anomalous Hall effect~(AHE)~\cite{Karplus1954,Nozieres1973,Crepieux2001,Sinitsyn2006,Inoue2006,Sinitsyn2007,Tanaka2008a,Kovalev2009,Nagaosa2009,Kovalev2010,Lowitzer2010} and the spin-orbit torques~\cite{Manchon2008,Manchon2009,Miron2010,Kurebayashi2014}.
 These phenomena have been studied enormously with fundamental as well as applicational interest.
 Among these, AMR, the change in electric conductivity upon varying the magnetization direction, was observed experimentally more than 150 years ago~\cite{Thomson1857} and is one of the most accessible physical quantities in experiments as well as AHE.

 The theoretical calculations of AMR were done mainly based on specific systems, such as the $3d$-transition metals, diluted magnetic semiconductors~(DMSs) and strongly spin-orbit coupled systems.
 For the $3d$-transition metals, the AMR is explained by using the two-current model~\cite{Smit1951,Campbell1970,kokado2012} and the \textit{ab initio} results showed good agreement with experiment~\cite{Banhart1995,Lowitzer2010}.
 The AMR depends on various factors including the $\bm{L} \cdot \bm{S}$-type SOC, the hybridization and the density of states of $s$- and $d$-electrons.
 For DMS ferromagnets such as (Ga, Mn)As, it is possible to calculate microscopically based on simple physical methods~\cite{Rushforth2007}.
 The AMR is determined mainly by the anisotropy of the lifetime induced by the combination of SOC with polarization of randomly distributed magnetic scatterers, rather than by that with polarization of conducting electrons resulting in an anisotropic band structure~\cite{Rushforth2007}.
 In this paper, we focus on the AMR of the strongly spin-orbit coupled systems, such as the interface between the ferromagnetic metal and the heavy metals and the magnetic semiconductors without inversion symmetry.
 It should be noted that the crystalline anisotropy can contribute to the AMR, but we here have focused on the noncrystalline AMR.

 In the context of the AMR of a strongly spin-orbit coupled system, the two-dimensional~(2D) Rashba ferromagnet was studied in two cases.
 In one case where the magnetization is made from randomly distributed magnetic impurities similar to DMS ferromagnets, a finite AMR due to the anisotropy of the lifetime was obtained by using the relaxation-time approximation~\cite{Vyborny2009}.
 In the other case that the exchange field is treated non-perturbatively, it was found that the AMR vanishes in the clean case~\cite{Kato2008} because the ladder type vertex corrections~(VCs) cancel out the bare-bubble contribution, as can be seen in AHE~\cite{Inoue2006}.
 The two results seem to contradict each other since the relaxation-time treatment is equivalent to a perturbation theory of the exchange field.
 This discrepancy is possibly explained by the contribution from the anisotropy of the band structure~\cite{McGuire1975,Rushforth2007}, which was not taken into account in the relaxation-time treatment.

 In order to reveal the microscopic origins of the AMR for simple metallic ferromagnets with strong SOC, we calculated an AMR of the three-dimensional~(3D) Dirac ferromagnet~\cite{Fujimoto2014}, an extension of the Stoner-type ferromagnet to the relativistic region in the previous work.
 The Dirac ferromagnet has two kinds of ferromagnetic order parameters in general: `magnetization' $\bm{M}$ and `spin' $\bm{S}$. 
 The AMR was found to be determined by the anisotropy of the group velocity resulting from the anisotropic band structure, in addition to the anisotropy of the lifetime.
 In the previous work, we calculated the diagonal conductivities without VCs.
 $\bm{M} = M \hat{z}$ and $\bm{S} = S \hat{z}$ are assumed, and the diagonal conductivity without VCs is denoted by $\sigma^{\mathrm{b}}_{A}$ ($A = \,\parallel, \perp$), where $\parallel$ ($\perp$) means the conductivity parallel (perpendicular) to the $\hat{z}$-direction.
 We found that the sign of AMR defined as $(\sigma_{\perp} - \sigma_{\parallel}) / (\sigma_{\perp} + \sigma_{\parallel})$ is opposite between the two typical cases; (i) $M > 0$, $S = 0$ and (ii) $M = 0$, $S > 0$, and the AMR magnitudes of (i) and (ii) are comparable and large (5 $\sim$ 25\%).
 This is because the deformation of the Fermi surfaces by $M$ is in the opposite way compared to that by $S$, and the deformations contribute to the anisotropy of the group velocity.
 For the coexistent case (iii) $M = S > 0$, there is no deformation of the Fermi surfaces, and then only the anisotropy of the damping determines the AMR, whose magnitude is smaller (0.1 $\sim$ 1\%).

 In this paper, we first evaluate the contributions of the ladder type VCs to $\sigma_{A}$ ($A = \,\parallel, \perp$), which are known to be important for the 2D Rashba ferromagnet~\cite{Kato2008}.
 For the Dirac ferromagnet, we find that they do not change AMR quantitatively since they almost equally increase $\sigma_{\perp}$ and $\sigma_{\parallel}$.
 We also show that there are two kinds of contributions: the renormalization of the lifetime and the additional spin-current contribution.
 The former contribution shows that the transport lifetime $\tau_{\mathrm{tr}}$ is different from the one-electron lifetime $\tau$, as is known in the electron gas system with the long-range impurity potential~\cite{Mahan2000} and in the 2D Dirac electron system with the short-range impurity potential~\cite{Sakai2014}.
 The latter contribution is due to the correlation between the electric and the spin currents through the impurity scattering.
 The Dirac ferromagnet has two kinds of spin currents; `magnetization'-current, $\bm{j}_{M}^{z}$, and `spin'-current, $\bm{j}_{S}^{z}$ (the superscript $z$ representing the component of the `magnetization' or `spin'), according to the two kinds of the ferromagnetic order parameters.
 We find that the additional spin-current contribution to $\sigma_{\perp}$ can be represented by using the correlations between the current and the `magnetization'-current, while the contribution to $\sigma_{\parallel}$ can be written by using the correlations between the electric current and the `spin'-current, since $\bm{j}_{M}^{z}$ ($\bm{j}_{S}^{z}$) flows only in the direction perpendicular (parallel) to the $\hat{z}$-direction.

 We also calculate the important contributions to AHE from the ladder type and skew-scattering type VCs.
 In the previous paper~\cite{Fujimoto2014}, we evaluated the transverse conductivity only without VCs which contains the intrinsic contribution to AHE and the part of the side-jump ones.
  We find that the skew-scattering contribution is proportional to $m c^2 / \Ni u$ and dominates AHE in the clean case when the chemical potential lies in the band~\cite{Nagaosa2009}, where $2 m c^2$ is the mass gap with $m$ being the mass of electron, the $c$ the speed of light, and $u$ is the impurity potential.
 This can be also seen in the spin Hall effect of the (non-magnetic) Dirac electron system~\cite{Fukazawa2017}.

\section{\label{sec:formulation}Formulation}
 Following the previous paper~\cite{Fujimoto2014}, we start with the $4 \times 4$ Hamiltonian,
\begin{align}
 \mathcal{H}_0 =  \hbar c \bm{k} \cdot \bm{\sigma} \rho_1 + mc^2 \rho_3 - \bm{M} \cdot \bm{\sigma} \rho_3 - \bm{S} \cdot \bm{\sigma}
\label{eq:Hamiltonian}
,\end{align}
where $\rho_i$ ($i = 1,2,3$) and $\sigma^j$ ($j = x,y,z$) are the Pauli matrices in particle-hole space and spin space, respectively.
 `Magnetization' $\bm{M}$ and  `spin' $\bm{S}$ are two kinds of ferromagnetic order parameters and assumed to be along the $\hat{z}$-direction, $\bm{M} = M \hat{z}$ and $\bm{S} = S \hat{z}$.
 In this paper, we consider $M < m c^2$ and $S < m c^2$, which corresponds to the non-topological~(trivial) phase.

 It may be worth pointing out the two kinds of the ferromagnetic order parameters of the Dirac ferromagnet; $\bm{M}$ and $\bm{S}$.
 In the literature of the spin-density functional theory, the model containing only $\bm{M}$ was first introduced by MacDonald and Vosko~\cite{MacDonald1979}, and the model containing only $\bm{S}$ was proposed by Ramana and Rajagopal~\cite{Ramana1979}.
 It is emphasized that the Dirac ferromagnet can be applied both to the relativistic case and to the case of a low-energy effective model of electrons in solids, where $c$ and $mc^2$ are replaced by the effective velocity and the energy gap, respectively.
 It is true that for the relativisitc case, $\bm{M}$ couples to the magnetic field $\bm{B}$ and describes the magnetic moment, while $\bm{S}$ stands for the spin but does not couple to any (electromagnetic) fields~\footnote{Considering the relativistic case in accelerated frames, the spin can couple to the mechanical rotation~\cite{Matsuo2011a}.}.
 In the case of an effective model, however, it is possible that there is a coupling between $\bm{S}$ and $\bm{B}$.
 From these, we treat $\bm{M}$ and $\bm{S}$ on an equal footing.
 Note that a Weyl semimetal~\cite{Burkov2011,Burkov2014a,Shitade2017} and a Dirac nodal semimetal~\cite{Burkov2011a,Xie2015,Yamakage2015} are described by the specific cases of the Dirac ferromagnet, $S > m c^2$ and $ M > m c^2$, respectively.
 We put $c = \hbar = 1$ and the volume of the system to unity hereafter.

The Green function for $\mathcal{H}_0$ is defined by $G^{(0)}_{\bm{k}} (\epsilon) = \{ \epsilon - \mathcal{H}_0 \}^{-1}$, and by using the Pauli matrices it is expressed as
\begin{align}
G^{(0)}_{\bm{k}} (\epsilon) = \frac{1}{D_{\bm{k}} (\epsilon)} \sum_{ \substack{ \mu = 0,1,2,3 \\ \nu = 0,x,y,z } } g^{(0)}_{\mu \nu} (\epsilon) \rho_{\mu} \sigma^{\nu}
.\end{align}
 Equation~(8) and Table I in Ref.~\cite{Fujimoto2014} gives the explicit forms of $D_{\bm{k}} (\epsilon)$ and $g^{(0)}_{\mu \nu} (\epsilon)$.

 As in the previous paper, we consider the randomly distributed impurity whose potential is the $\delta$-function type.
 In order to take the skew-scatting contribution to AHE into account according to the Ward-Takahashi identity, the self-energy due to the potential should be considered within the self-consistent $T$-matrix approximation.
 However, by assuming the clean case, $\Ni \ll 1$ and $u \sum_{\eta = \pm 1} \nu_{0,0}^{\eta} (\mu) \ll 1$, where $\Ni$ and $u$ are the impurity concentration and potential, and $\nu_{0,0}^{\eta} (\mu)$ is $\eta$-band's density of states at the Fermi level defined by (A4) in Ref.~\cite{Fujimoto2014}, the self-energy is approximated to that within the Born approximation, and it is given by Eq.~(13) in Ref.~\cite{Fujimoto2014}.
 The renormalized retarded/advanced Green function $G_{\bm{k}}^{\R/\A} (\epsilon)$ is given by $\{ G_{\bm{k}}^{\R/\A} (\epsilon) \}^{-1} = \{ G^{(0)}_{\bm{k}} (\epsilon+(-)\zi 0) \}^{-1} - \varSigma^{\R/\A} (\epsilon)$.

 We now calculate the conductivity tensor with VCs.
 We evaluate the conductivity from the retarded current-current correlation function divided by $\zi \omega$, by taking the limit, $\omega \to 0$, where $\omega$ is the frequency of the external electric field.
 The temperature is assumed to be absolute zero.
 Then, the diagonal conductivity in the clean case is expressed as
\begin{align}
\sigma_{i i}^{}
	& =
		\frac{e^2}{2 \pi} \sum_{\bk}
		\tr \left[
			v_{i} G_{\bk}^{\R} (\epsilon)
				\tilde{\varLambda}_{1,i} G_{\bm{k}}^{\A} (\epsilon)
		\right] \bigl. \Bigr|_{\epsilon=\mu}
\label{eq:AMR-wVC}
\end{align}
with $i = x, y, z$, where we neglected VCs which consist only of the retarded (or advanced) Green functions because they contribute in the higher orders with respect to $\Ni u^2$.
\begin{figure}[b!]
\centering
\includegraphics[page=1,width=0.4\linewidth]{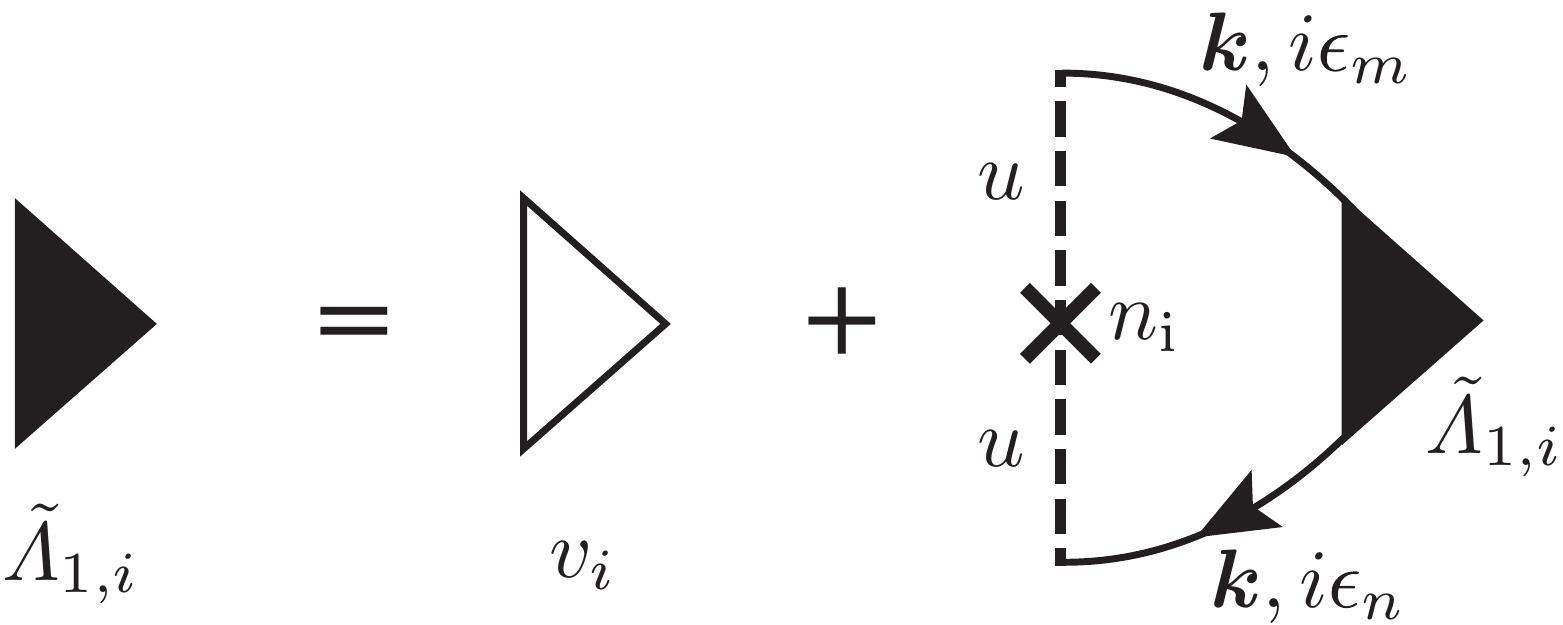}
\caption{\label{fig:velocity_vc}
 The Feynman diagram of the velocity vertex.
 The filled (unfilled) triangle describes the full (bare) velocity vertex.
 The solid line represents the Matsubara Green function.
 The dotted line and the cross symbol denote the impurity potential and concentration, respectively.
 Note that \cref{eq:varLambda_sj} is obtained from the equation given by this diagram after taking the analytic continuation $\zi \epsilon_m \to \epsilon + \omega + \zi 0$ and $\zi \epsilon_n \to \epsilon - \zi 0$ and taking the limit $\omega \to 0$.
}
\end{figure}
 As shown diagrammatically in \cref{fig:velocity_vc}, the full velocity vertex $\tilde{\varLambda}_{1,i}$ is given as
\begin{align}
\tilde{\varLambda}_{1,i} (\epsilon)
	& =
	v_i
	+ \Ni u^2 \sum_{\bm{k}'}
		G_{\bm{k}'}^{\R} (\epsilon)
		\tilde{\varLambda}_{1,i} (\epsilon)
		G_{\bm{k}'}^{\A} (\epsilon)
\label{eq:varLambda_sj}
\end{align}
with the bare velocity vertex defined by
\begin{align}
v_i
	& = \frac{\partial \mathcal{H}_{\bk}}{\partial k_i}
	= \rho_1 \sigma^i
\label{eq:def_charge_current_velocity}
.\end{align}
\begin{figure*}[bt!]
\centering
\includegraphics[page=2,width=0.9\linewidth]{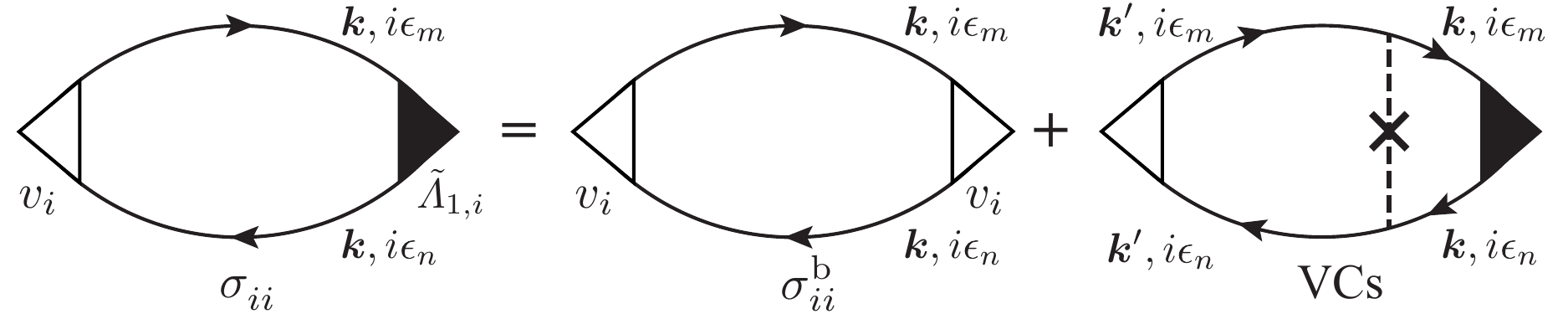}
\caption{\label{fig:diagonal}
 The diagrammatic expressions of the diagonal conductivity and the two decomposed contributions from the bare bubble and the VCs.
 The lines and symbols are defined in the caption of \cref{fig:velocity_vc}.
 These diagrams are defined in the Matsubara formalism as well as \cref{fig:velocity_vc}, and \cref{eq:AMR-wVC_decomposed} is obtained after rewriting the Matsubara summation of $\zi \epsilon_n$ using the contour integral, taking the analytic continuation $\zi \epsilon_m - \zi \epsilon_n \to \omega + \zi 0$, and leaving the $\omega$-linear term which includes both retarded and advanced Green functions.
}
\end{figure*}
\begin{figure*}[bt!]
\centering
\includegraphics[page=3,width=0.9\linewidth]{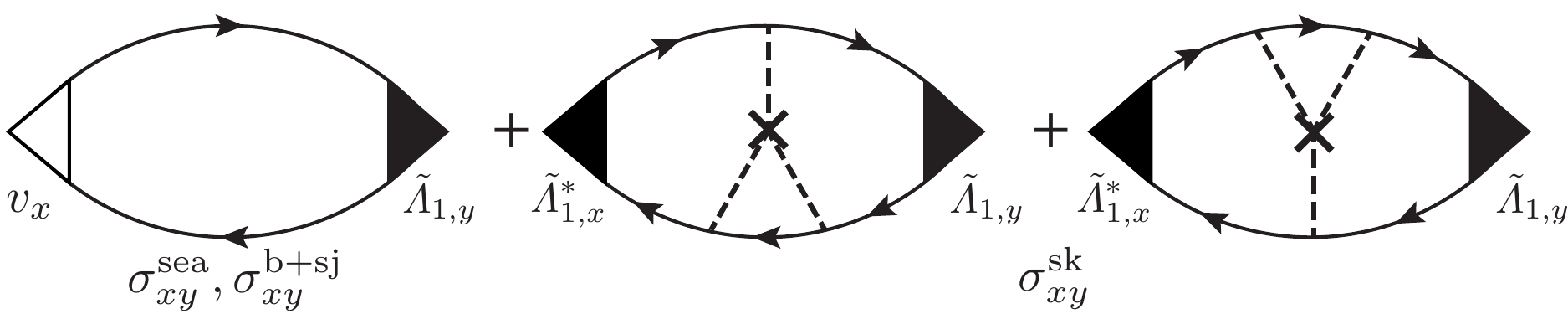}
\caption{\label{fig:off-diagonal}
 The diagrammatic expressions of the intrinsic, side-jump and skew-scattering contributions to the AHE.
 The skew-scattering contribution can be given as the two Feynman diagrams for the case $\Ni \ll 1$ and $u \sum_{\eta} \nu^{\eta}_{0,0} (\mu) \ll 1$.
 \Cref{eq:AHE-b+sj,eq:AHE-sk} are obtained through the procedures noted in the caption of \cref{fig:diagonal}.
}
\end{figure*}
 \Cref{eq:AMR-wVC} can be decomposed as
\begin{align}
\sigma_{i i}^{}
	& 	 = \sigma_{i i}^{\mathrm{b}}
		+ \frac{e^2}{2 \pi} \sum_{\bk}
		\tr \left[
			v_{i} G_{\bk}^{\R} (\epsilon)
				\varLambda_{1,i} G_{\bm{k}}^{\A} (\epsilon)
		\right] \bigl. \Bigr|_{\epsilon=\mu}
\label{eq:AMR-wVC_decomposed}
,\end{align}
where the first term is the bare-bubble contribution $\sigma_{i i}^{\mathrm{b}}$ identical to Eq.~(28) in Ref.~\cite{Fujimoto2014}, and the second term is contributions of the VCs, $\varLambda_{1,i} = \tilde{\varLambda}_{1,i} - \rho_1 \sigma^{i}$ (see \cref{fig:diagonal}).

 The off-diagonal conductivity is obtained as $\sigma_{yx} = - \sigma_{xy}$ with
\begin{align}
\sigma_{xy}
	& = \sigma_{xy}^{\mathrm{sea}}
		 + \sigma_{xy}^{\mathrm{b+sj}}
		 + \sigma_{xy}^{\mathrm{sk}}
\label{eq:sigma_xy}
,\end{align}
and $\sigma_{iz} = \sigma_{zi} = 0$ with $i = x, y$ as expected from the symmetry of the configurations.
 Here, $\sigma_{xy}^{\mathrm{sea}}$ is the contribution from the states below the Fermi level (Fermi sea) and given as Eqs.~(30) in Ref.~\cite{Fujimoto2014}.
 To be precise, there are the VCs to the Fermi-sea contribution, but they are higher order with respect to $\Ni u^2$ and hence neglected.
\begin{align}
\sigma_{xy}^{\mathrm{b+sj}}
	& = \frac{e^2}{4 \pi} \sum_{\bk}
		\mathrm{tr} \left[
			v_{x} G_{\bm{k}}^{\R}
				\tilde{\varLambda}_{1,y} G_{\bk}^{\A}
			- v_{y} G_{\bm{k}}^{\R}
				\tilde{\varLambda}_{1,x} G_{\bk}^{\A}
		\right] \bigl. \Bigr|_{\epsilon=\mu}
\label{eq:AHE-b+sj}
\end{align}
is the contribution from the bare-bubble contribution from the states at the Fermi level~(Fermi surface) and the ladder type VCs.
 By using $\tilde{\varLambda}_{1,i} = \rho_1 \sigma^{i} + \varLambda_{1,i}$, this term is decomposed as
\begin{align}
\sigma_{xy}^{\mathrm{b+sj}}
	& =\sigma_{xy}^{\mathrm{b}}
	+ \frac{e^2}{4 \pi} \sum_{\bk}
		\mathrm{tr} \left[
			v_{x} G_{\bm{k}}^{\R}
				\varLambda_{1,y} G_{\bk}^{\A}
			- v_{y} G_{\bm{k}}^{\R}
				\varLambda_{1,x} G_{\bk}^{\A}
		\right] \bigl. \Bigr|_{\epsilon=\mu}
\label{eq:AHE-sj}
,\end{align}
and the bare-bubble contribution $\sigma_{xy}^{\mathrm{b}}$ corresponds to Eq.~(29) in Ref.~\cite{Fujimoto2014}.
 The last term of \cref{eq:sigma_xy} is contribution from the skew-scattering type VC given by
\begin{align}
\sigma_{x y}^{\mathrm{sk}}
	& = \frac{e^2 \Ni u^3}{4 \pi} \sum_{\bk, \bk', \bk''}
		\mathrm{tr} \left[
			\tilde{\varLambda}^{*}_{1,x}
			 G^{\R}_{\bk}
			 G^{\R}_{\bk''}
			 G^{\R}_{\bk'}
			\tilde{\varLambda}_{1,y}
			 G^{\A}_{\bk'}
			 G^{\A}_{\bk}
\right. \notag \\ & \hspace{1em} \left.
			+ \tilde{\varLambda}^{*}_{1,x}
			 G^{\R}_{\bk}
			 G^{\R}_{\bk'}
			\tilde{\varLambda}_{1,y}
			 G^{\A}_{\bk'}
			 G^{\A}_{\bk''}
			 G^{\A}_{\bk}
		- ( \R \leftrightarrow \A )
		\right]
		\Bigr|_{\epsilon = \mu}
\label{eq:AHE-sk}
,\end{align}
where $\tilde{\varLambda}^{*}_{1,i}$ ($i = x,y$) is defined by interchanging $\R$ and $\A$ in $\tilde{\varLambda}_{1,i}$.
 \Cref{fig:off-diagonal} depicts the Feynman diagrams of the off-diagonal conductivity.

 The detailed calculation of \cref{eq:varLambda_sj} is presented in \cref{apx:vc}, and calculations of \cref{eq:AMR-wVC,eq:AHE-sj,eq:AHE-sk} are given in \cref{apx:conductivity}.

\section{\label{sec:result-discussion}Results and Discussion}

\subsection{\label{sec:sub:AMR}Diagonal conductivity and AMR}
 The diagonal conductivities, $\sigma_{ii}$ ($i = x, y, z$), are rewritten as that for perpendicular ($\sigma_{\perp} = \sigma_{xx} = \sigma_{yy}$) and that for parallel ($\sigma_{\parallel} = \sigma_{zz}$) configurations.
 First of all, we rewrite the diagonal conductivity without VCs, $\sigma^{\rm b}_{A}$ ($A = \perp, \parallel$) [Eq.~(44) with Eqs.~(45) and (46) in Ref.~\cite{Fujimoto2014}], in terms of the correlation function between dimensionless operators $P$ and $Q$, $\bbracket{P; Q}$,
\begin{align}
\bbracket{P; Q}
	& = \frac{1}{4} \Ni u^2 \sum_{\bk} \tr \left[
		G^{\R}_{\bk} (\epsilon) \hat{P} G^{\A}_{\bk} (\epsilon) \hat{Q}
	\right] \Big|_{\epsilon = \mu}
\label{eq:correlation_def}
,\end{align}
where $P (Q) = \int \dd{\br} \psi^{\dagger} (\br) \hat{P} (\hat{Q}) \psi^{} (\br)$ is dimensionless operators with the annihilation~(creation) operator of the field $\psi^{(\dagger)} (\br)$.
 Here, $\bbracket{P; Q}$ denotes the correlation evaluated only from the bare-bubble diagram and do not include any VCs.
 By using this representation, $\sigma^{\rm b}_{A}$ are expressed simply as
\begin{align}
\sigma^{\mathrm{b}}_{\perp}
	& =
	\frac{2 e^2}{\pi \Ni u^2}
		\bbracket{j_x; j_x}
\label{eq:result_sigma_perp_woVC}
, \\
\sigma^{\mathrm{b}}_{\parallel}
	& = 
	\frac{2 e^2}{\pi \Ni u^2}
		\bbracket{j_z; j_z}
\label{eq:result_sigma_para_woVC}
,\end{align}
where $j_i$ ($i = x, z$) is the particle-current with the velocity given by \cref{eq:def_charge_current_velocity}, and the explicit forms of $\bbracket{j_x; j_x}$ and $\bbracket{j_z; j_z}$ are shown by \cref{eq:jx-jx,eq:jz-jz} in the leading order with respect to $\Ni u^2$.

 The diagonal conductivities with the ladder type VCs [\cref{eq:AMR-wVC}] are expressed as
\begin{align}
\sigma_{A}^{}
	& = \tilde{\sigma}_{A}^{\rm b}
		+ \sigma^{\mathrm{add}}_{A}
\qquad (A = \perp, \parallel)
\label{eq:result_sigma}
,\end{align}
where the first terms are given as the direct correlation functions renormalized by the ladder type VCs,
\begin{align}
\tilde{\sigma}_{\perp}^{\rm b}
	& =
		\frac{ 1 - \bbracket{j_{M, x}^{z}; j_{M, x}^{z}} }{ \mathcal{D}_{\perp} }
		\sigma^{\mathrm{b}}_{\perp}
\label{eq:result_sigma_perp}
, \\
\tilde{\sigma}_{\parallel}^{\rm b}
	& =
		\frac{ 1 - \bbracket{j_{S,z}^{z}; j_{S,z}^{z}} }{ \mathcal{D}_{\parallel} }
		\sigma^{\mathrm{b}}_{\parallel}
\label{eq:result_sigma_para}
,\end{align}
$\mathcal{D}_{A}$ ($A = \perp, \parallel$) are given by
\begin{align}
\mathcal{D}_{\perp}
	& = \left( 1 - \bbracket{j_x; j_x} \right) \left( 1 - \bbracket{j^{z}_{M,x}; j^{z}_{M,x}} \right)
		 - \bbracket{j_x; j^{z}_{M,x}}^2
\label{eq:def_D_perp}
, \\
\mathcal{D}_{\parallel}
	& = \left( 1 - \bbracket{j_z; j_z} \right) \left( 1 - \bbracket{j^{z}_{S,z}; j^{z}_{S,z}} \right)
		- \bbracket{j_z; j^{z}_{S,z}}^2
\label{eq:def_D_para}
,\end{align}
and $\bbracket{j_{M, x}^{z}; j_{M, x}^{z}}$, $\bbracket{j_x; j^{z}_{M,x}}$, $\bbracket{j_{S,z}^{z}; j_{S,z}^{z}}$, and $\bbracket{j_z;  j^{z}_{S,z}}$ are given by \cref{eq:jMx-jMx,eq:jx-jMx,eq:jz-jz,eq:jz-jSz}.
 Here, $j_{M,i}^{\alpha}$ and $j_{S,i}^{\alpha}$ ($\alpha, i = x, y, z$) are `magnetization'-, and `spin'-currents with their velocities, respectively, given by
\begin{align}
v^{\alpha}_{M, i}
	& = \frac{1}{2} \{ \rho_3 \sigma^\alpha, \rho_1 \sigma^i \}
	  = \sum_{\beta = x, y, z} \varepsilon_{i \alpha \beta} \rho_2 \sigma^{\beta}
\label{eq:def_magnetization_current_velocity}
, \\
v^{\alpha}_{S, i}
	& = \frac{1}{2} \{ \sigma^\alpha, \rho_1 \sigma^i \}
	  = \rho_1 \delta_{i, \alpha}
\label{eq:def_spin_current_velocity}
,\end{align}
where $\{\hat{P}, \hat{Q}\} = \hat{P} \hat{Q} + \hat{Q} \hat{P}$ is the anticommutater.
 The second terms of \cref{eq:result_sigma} are additional contributions by considering the ladder type VCs,
\begin{align}
\sigma^{\mathrm{add}}_{\perp}
	& =
	\frac{2 e^2}{\pi \Ni u^2}
	\frac{\bbracket{j_x; j_{M,x}^{z}} \bbracket{j_{M,x}^{z}; j_x}}{\mathcal{D}_{\perp}}
\label{eq:result_sigma_add_perp}
, \\
\sigma^{\mathrm{add}}_{\parallel}
	& =
	\frac{2 e^2}{\pi \Ni u^2}
	\frac{\bbracket{j_z; j_{S,z}^{z}} \bbracket{j_{S,z}^{z}; j_z}}{\mathcal{D}_{\parallel}}
\label{eq:result_sigma_add_para}
,\end{align}
which are given as the correlation functions of the electric-current with `magnetization'- and `spin'-current.

\begin{figure*}[hbtp]
\centering
\includegraphics[width=0.9\linewidth]{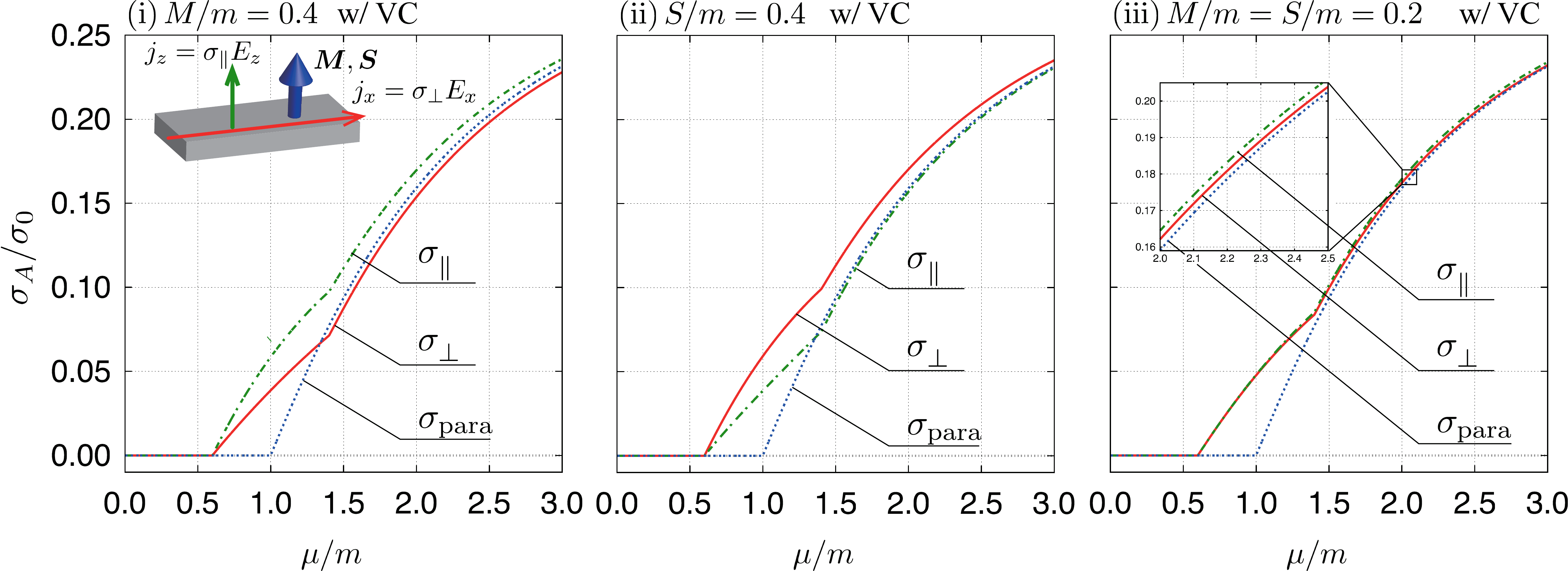}
\caption{\label{fig:AMR-wVC}(Color online)~%
 The diagonal conductivities with the ladder type VCs, $\sigma_{A}$ ($A = \perp, \parallel$), for the typical three cases (i)-(iii) as functions of the chemical potential $\mu$.
 The inset of (i) describes the configuration of the directions in which the current flows and the vector of the `magnetization' and/or `spin'.
 They are normalized by $\sigma_0 = e^2 m^2 c^3 / \hbar^2 \gamma$ with $\gamma = \Ni u^2 m^2 c / \hbar^3$.
 In the region of $\mu / m < 0.6$, the chemical potential lies in the gap.
 In the regions of $0.6 < \mu < 1.4$ and $\mu > 1.4$, the system has one and two Fermi surfaces, respectively.
 Those in the paramagnetic state ($M = S = 0$) is also shown for comparison.
}
\includegraphics[width=0.9\linewidth]{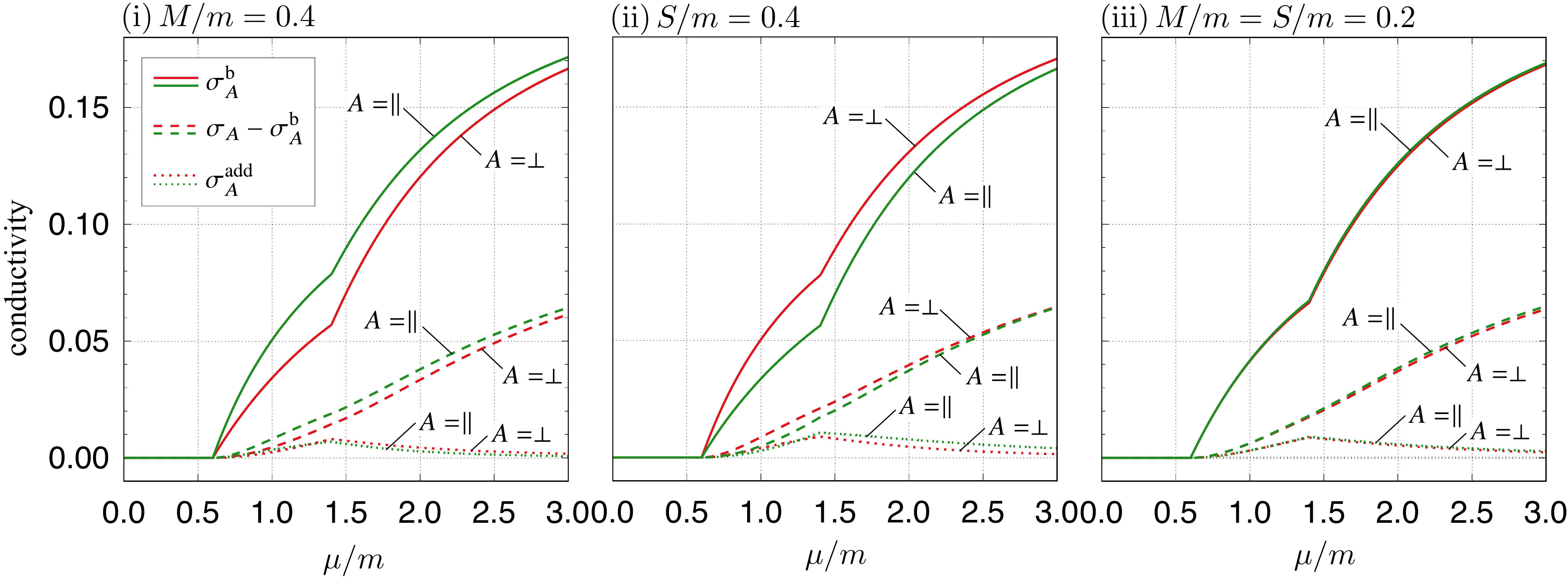}
\caption{\label{fig:AMR-eVC}(Color online)~%
 The $\mu$-dependences of the the bare-bubble and the VC contributions to the diagonal conductivities for the typical three cases (i)-(iii).
 Note that the VC contributions are given by $\sigma^{}_{A} - \sigma^{\rm b}_{A}$.
 The additional contribution $\sigma^{\rm add}_{A}$, which is included by the VC contributions, is also shown for comparison.
 They are normalized by $\sigma_0$ given in the caption of \cref{fig:AMR-wVC}.
}
\end{figure*}
 \Cref{fig:AMR-wVC} shows $\sigma_{A}$ ($A = \perp, \parallel$) as functions of the chemical potential $\mu$ for the typical three cases; (i) $M > 0$, $S = 0$, (ii) $M = 0$, $S > 0$, and (iii) $M = S > 0$, together with the one in the paramagnetic state ($M = S = 0$).
 \Cref{fig:AMR-eVC} depicts the individual contributions to $\sigma_{A}$ as functions of $\mu$, where the contribution from the VCs, $\sigma_{A} - \sigma^{\rm b}_{A} = (\tilde{\sigma}^{\rm b}_{A} - \sigma^{\rm b}_{A}) + \sigma^{\rm add}_{A}$, contains both the renormalization of $\sigma^{\rm b}_{A}$ and the additional spin-current contribution.
 From \cref{fig:AMR-wVC,fig:AMR-eVC}, the VCs increase the conductivities, and $\sigma_{A}$ in \cref{fig:AMR-wVC} is about $1.3$ times larger than $\sigma^{\rm b}_{A}$ in \cref{fig:AMR-eVC} for all three cases.
 However, qualitative dependences do not change even by considering the VCs, and hence the AMR ratio with ladder type VCs defined by
\begin{align}
R
	& =  \frac{\sigma_{\perp} - \sigma_{\parallel} }{ \sigma_{\perp} + \sigma_{\parallel} }
\end{align}
is quantitatively same as that without VCs, $R \simeq R_{\mathrm{b}}$, where
\begin{align}
R_{\mathrm{b}}
	& =  \frac{\sigma^{\rm b}_{\perp} - \sigma^{\rm b}_{\parallel} }{ \sigma^{\rm b}_{\perp} + \sigma^{\rm b}_{\parallel} }
.\end{align}
 From the above, we conclude that we do not need to consider the VCs for the AMR in the Dirac ferromagnet.
 This is different from the case of the 2D Rashba ferromagnet~\cite{Kato2008}.

\begin{figure}[hbtp]
\centering
\includegraphics[width=0.4\linewidth]{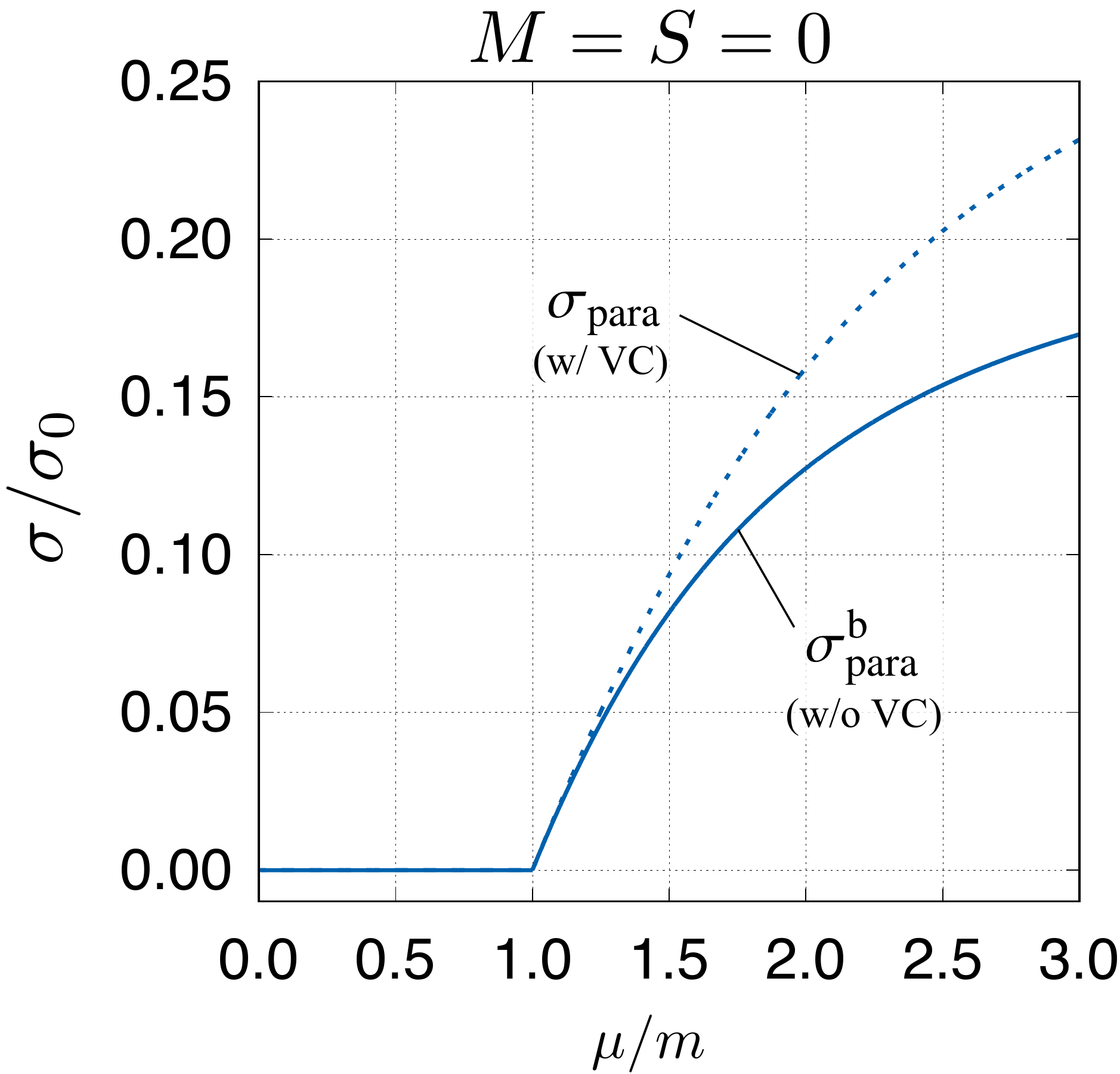}
\caption{\label{fig:conductivity_M=S=0}(Color online)~%
 The diagonal conductivities for $M = S = 0$ with and without the ladder type VCs.
 The VCs increase the conductivity, which can be understood as the fact that the forward scattering is dominant in the impurity scattering for the diagonal conductivity.
}
\end{figure}
 The additional contributions, $\sigma^{\mathrm{add}}_{A}$ ($A = \perp, \parallel$), can be understood as an effect that the impurity scatterings interchange between the particle current and spin current.
 Since the Dirac ferromagnet has two types of the order parameters, `magnetization' and `spin', there are the two corresponding spin currents, `magnetization'- and `spin'-currents, and they flow only in the specific directions, respectively, as in \cref{eq:def_magnetization_current_velocity,eq:def_spin_current_velocity}.
 Hence, the charge-current in the $\hat{x}$- or $\hat{y}$-directions includes only the contribution from the `magnetization'-current, while the `spin'-current contributes only to the charge-current in the $\hat{z}$-direction.
 Note that $\sigma_{A}^{\rm add}$ is not dominant contributions of the VCs, as can be seen in \cref{fig:AMR-eVC}.

 We demonstrate that the renormalization [\cref{eq:result_sigma_perp,eq:result_sigma_para}] can be understood as the similar change of the lifetime into the transport lifetime for the case of $M = S = 0$ (See \cref{fig:conductivity_M=S=0}).
 For the simple case, the Green function is given as
\begin{align}
G^{\R}_{\bk} (\mu) \big|_{M=S=0}
	& = \sum_{\eta} \frac{ \ket{\varphi_{\bk}} \bra{\varphi_{\bk}} }{\mu - \eta \epsilon_k + \zi \Gamma }
,\end{align}
where $\pm \epsilon_k = \pm \sqrt{ k^2 + m^2 }$ are the eigen energies, $\ket{\varphi_{\bk}} \bra{\varphi_{\bk}} = (\mu + \rho_1 \bk \cdot \bm{\sigma} + \rho_3 m) / 2 \mu$ and
\begin{align}
\Gamma
	& = \frac{\pi}{2} \Ni u^2 \nu (\mu) \left( 1 + \frac{m^2}{\mu^2} \right)
\label{eq:damping_MS=0}
\end{align}
is the damping of electron.
 In this case, the two spin currents have no correlations with the charge current in the leading order with respect to $\Ni$~\cite{Fukazawa2017}, and the ladder type VCs is reduced to
\begin{align}
\tilde{\varLambda}_{1,x}
	& = \rho_1 \sigma^x
		+ \Ni u^2 \sum_{\bk} G^{\R}_{\bk} (\mu) \tilde{\varLambda}_{1,x} G^{\A}_{\bk} (\mu)
.\end{align}
 This can be solved by presuming $\tilde{\varLambda}_{1,x} = U \rho_1 \sigma^x$, multiplying the both sides by $\rho_1 \sigma^x / 4$, and taking the trace,
\begin{align}
U
	& = \left( 1 - \frac{\Ni u^2}{4} \sum_{\bk} \tr [ G^{\R}_{\bk} (\mu) \rho_1 \sigma^x G^{\A}_{\bk} (\mu) \rho_1 \sigma^x ] \right)^{-1}
	+ \Ord{\Ni}
\label{eq:correspondence_MS=0}
\\ &
	= \frac{ \Gamma }{ \Gamma - \Gamma' }
	+ \Ord{\Ni}
,\end{align}
where $\Gamma'$ is given as
\begin{align}
\Gamma'
	& = \frac{\pi}{2} \Ni u^2 \nu (\mu) \frac{1}{3} \left( 1 - \frac{m^2}{\mu^2} \right)
\label{eq:damping'_MS=0}
.\end{align}
 Hence, the conductivities with and without the ladder type VCs are obtained as
\begin{align}
\sigma_{\rm para}
	& = \frac{2 e^2}{\pi \Ni u^2} \frac{\Gamma' / \Gamma}{1 - U}
	= \frac{2 e^2}{\pi \Ni u^2} \frac{\Gamma'}{\Gamma - \Gamma'}
, \\
\sigma_{\rm para}^{\rm b}
	& = \frac{2 e^2}{\pi \Ni u^2} \frac{\Gamma'}{\Gamma}
.\end{align}
 From these, the one-electron lifetime ($\tau = \hbar / \Gamma$) is changed into the transport lifetime ($\tau_{\rm tr}^{-1} = \tau^{-1} - \Gamma'/\hbar$) by considering the ladder type VCs.
 For the cases of $M \neq 0$ and/or $S \neq 0$, the lifetime depends on the spin, and the charge current has the correlations with the two kinds of the spin currents.
 The coefficient of $\sigma_{\perp}^{\rm b}$ [\cref{eq:result_sigma_perp}] is rewritten as
\begin{align}
\frac{ 1 - \bbracket{j_{M, x}^{z}; j_{M, x}^{z}} }{ \mathcal{D}_{\perp} }
	& = \left( 1 - \bbracket{j_x; j_x} - \frac{ \bbracket{j_x; j^{z}_{M,x}} \bbracket{j^{z}_{M,x}; j_x} }{ 1 - \bbracket{j^{z}_{M,x}; j^{z}_{M,x}} } \right)^{-1}
,\end{align}
where the first two terms can be regarded as the correspondences with \cref{eq:correspondence_MS=0}, but the last term is obtained only in the cases $M \neq 0$ or $S \neq 0$.

 We can see that the forward scattering is dominant over the impurity scatterings for the transport lifetime as other systems~\cite{Mahan2000,Vyborny2009,Sakai2014}.
 The one-electron lifetime is expressed as $\tau^{-1} \propto \int \dd{\Omega_p} W_{\bk, \bp}$, and the transport lifetime is written as $\tau_{\mathrm{tr}}^{-1} \propto \int \dd{\Omega_p} ( 1 - \cos \theta_{p}) W_{\bk, \bp}$, where $\bk$ and $\bp$ are the wavevectors on the Fermi surface, $W_{\bk,\bp}$ is the scattering amplitude between the wavevectors, $\int \dd{\Omega_p}$ is the integral of the solid angle of $\bp$, and $\cos \theta_p = \bk \cdot \bp / |\bk| |\bp| $.
 Here, we focus on the following systems with the non-magnetic impurity potential which has the $\delta$-function type.
 For the free electron gas, the scattering amplitude $W_{\bk,\bp}$ is independent of $\theta_p$, and hence the transport lifetime coincides the one-electron lifetime, $\tau_{\rm tr} = \tau$~\cite{Mahan2000}.
 On the other hand, for the spin-momentum locked systems such as the 2D massive Dirac system~\cite{Sakai2014}, the 2D Rashba system~\cite{Vyborny2009}, and the (non-magnetic) 3D Dirac electron system, $W_{\bk,\bp}$ depends on $\theta_p$, because the scattering amplitude is written as $W_{\bk,\bp} \propto | \braket{\varphi_{\bp}|\varphi_{\bk}} |^2$, and $\ket{\varphi_{\bk}}$ is in the spinor form.
 Therefore, the transport lifetime is different from the one-electron lifetime in these systems.
\subsection{\label{sec:sub:AHE}Off-diagonal conductivity (AHE)}
 We rewrite the off-diagonal conductivity without VCs, which is identical to Eq.~(52) in Ref.~\cite{Fujimoto2014}, as
\begin{align}
\sigma^{\mathrm{b}}_{xy}
	& = - \frac{ 2 e^2 }{ \pi \Ni u^2 } \bbracketsurf{j_x; j_y}
\label{eq:result_AHE_woVC}
,\end{align}
where $\bbracketsurf{j_x; j_y}$ is the Fermi-surface contribution to the correlation between $j_x$ and $j_y$ evaluated from the bare-bubble diagram and the explicit form is given by \cref{eq:jx-jy}.
 Here, $\bbracketsurf{P; Q}$ is defined by \cref{eq:correlation_def} as same as $\bbracket{P; Q}$, but we distinguish $\bbracketsurf{P; Q}$ from $\bbracket{P; Q}$ because off-diagonal correlation functions include the Fermi-sea contribution in general.
 In fact, $\bbracket{j_x; j_y} = \bbracketsurf{j_x; j_y} + (\pi \Ni u^2 / 2 e^2) \sigma_{xy}^{\mathrm{sea}}$ when evaluated without VCs, while $\bbracket{j_x; j_x} = \bbracketsurf{j_x; j_x}$.

 $\sigma_{x y}^{\rm b+sj}$ can be expressed similar to the diagonal conductivity as
\begin{align}
\sigma_{x y}^{\rm b+sj}
	& = \tilde{\sigma}^{\mathrm{b}}_{xy} + \sigma^{\mathrm{add}}_{xy}
\label{eq:result_AHE_b+sj}
,\end{align}
where the first term is given as
\begin{align}
\tilde{\sigma}^{\mathrm{b}}_{xy}
	& =
		\left( \frac{ 1 - \bbracket{j^{z}_{M,x}; j^{z}_{M,x}} }{ \mathcal{D}_{\perp} } \right)^2
		\sigma_{x y}^{\mathrm{b}}
\label{eq:result_AHE_b_tilde}
,\end{align}
which corresponds to $\sigma^{\mathrm{b}}_{xy}$ renormalized by the ladder type VCs, and
\begin{align}
\sigma^{\mathrm{add}}_{xy}
	& =
	\frac{2 e^2}{ \pi \Ni u^2 }
		\Biggl\{
			\Big( \bbracketld{j_x; j^{z}_{M,x}} \Big)^2
				\bbracketsurf{j^{z}_{M,x}; j^{z}_{M,y}}
			+ 2 \frac{ 1 - \bbracket{j^{z}_{M,x}; j^{z}_{M,x}} }{ \mathcal{D}_{\perp} }
			\bbracketld{j_x; j^{z}_{M,x}}
				\bbracketsurf{j^{z}_{M, x}; j_y}
		\Biggr\}
\label{eq:result_AHE_additional}
\end{align}
is the additional contribution by considering the ladder type VCs.
 Here, we introduced $\bbracketld{P; Q}$ as the correlation function evaluated within the ladder type VCs.
 For $P = j_x$ and $Q = j^{z}_{M,x}$, it is given by
\begin{align}
\bbracketld{j_x; j^{z}_{M,x}}
	& = \frac{1}{4} \Ni u^2 \sum_{\bk} \tr \left[ \rho_1 \sigma^x G^{\R}_{\bk} (\epsilon) \tilde{\varLambda}_{2,x} G^{\A}_{\bk} (\epsilon) \right] \Bigl|_{\epsilon = \mu}
\notag \\ &
	= \frac{1}{\mathcal{D}_{\perp}} \bbracket{j_x; j^{z}_{M,x}}
,\end{align}
and $\bbracketld{j_x; j^{z}_{M,x}} = \bbracketld{j^{z}_{M,x}; j_x} = - \bbracketld{j_y; j^{z}_{M,y}} = - \bbracketld{j^{z}_{M,y}; j_y}$ up to $\Ord{\Ni^0}$.
 (See also \cref{apx:transverse_ladder}.)
 \Cref{eq:result_AHE_additional} contains all the correlations through the `magnetization'-current assisted by the impurity scatterings.
 Note that there is no contribution of `spin'-current, because it does not flow in $xy$-plane.

 The skew-scattering type VC up to the leading order of $\Ni$ is obtained as
\begin{align}
\sigma^{\mathrm{sk}}_{x y}
 & = - \frac{4 e^2}{\pi \Ni u}
	\Biggl[
		\left\{
			\Bigl( \bbracketld{j_x; j_x} \Bigr)^2
			+ \Bigl( \bbracketld{j_x; j^{z}_{M,x}} \Bigr)^2
		\right\} \tilde{\gamma}_{0z}
		+ 2 \bbracketld{j_x; j^{z}_{M,x}} \bbracketld{j_x; j_x} \tilde{\gamma}_{30}
	 \Biggr]
\label{eq:result_AHE_skew}
,\end{align}
where $\tilde{\gamma}_{0z} = \gamma_{0z} / \Ni u^2$ and $\tilde{\gamma}_{30} =\gamma_{30} / \Ni u^2$ are $\Ord{\Ni^0}$ with $\gamma_{0z}$ and $\gamma_{30}$ being the damping constants, $\Im \varSigma^{\R} (\mu) = - \gamma_{\mu \nu} \rho_{\mu} \sigma^{\nu}$ [Eqs.~(A5) in Ref.~\cite{Fujimoto2014}].
 $\bbracketld{j_x; j_x} = \bbracketld{j_y; j_y}$ is given as
\begin{align}
\bbracketld{j_x; j_x}
	& = \frac{\pi \Ni u^2}{2 e^2} \sigma_{\perp}
.\end{align}

\begin{figure*}[hbtp]
\centering
\includegraphics[width=0.9\linewidth]{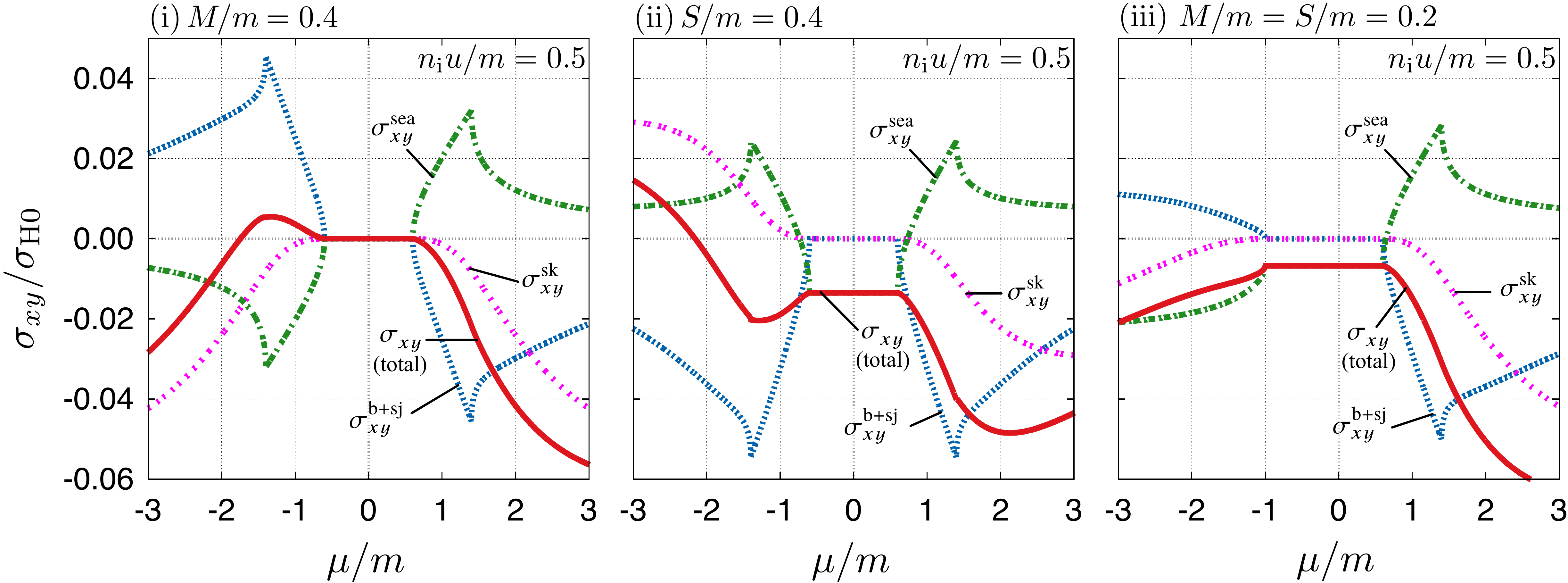}
	\caption{\label{fig:AHE-wVC}(Color online)~%
 The chemical potential dependences of the off-diagonal (Hall) conductivities for the three typical cases.
 The total Hall conductivities are no longer even/odd functions of $\mu$ because of the skew-scattering contribution $\sigma_{xy}^{\mathrm{sk}}$.
 Here, $\sigma_{xy}^{\mathrm{sk}}$ is shown for $\Ni u / m = 0.5$.
}
\end{figure*}

 In \cref{fig:AHE-wVC}, the off-diagonal conductivities in the typical three cases are shown as functions of the chemical potential, where the skew-scattering contribution is plotted for $\Ni u / m = 0.5$.
 We can see from \cref{fig:AHE-wVC} that the total Hall conductivities are neither even nor odd functions of the chemical potential in the three cases.
 From the symmetry consideration as discussed in Appendix~E of Ref.~\cite{Fujimoto2014}, we find the following relations,
\begin{align}
\sigma_{xy}^{\rm b+sj} (\mu, M, S)
	& = - \sigma_{xy}^{\rm b+sj} (-\mu, M, -S)
,\end{align}
which is the same symmetry as $\sigma^{\rm b}_{xy} (\mu, M, S)$ and $\sigma^{\rm sea}_{xy} (\mu, M, S)$, while the skew-scattering contribution has the different symmetry,
\begin{align}
\sigma^{\rm sk}_{xy} (\mu, M, S)
	& = \sigma^{\rm sk}_{xy} (-\mu, M, -S)
.\end{align}
 This difference can be explained by the number of the Green functions; $\sigma_{xy}^{\rm b+sj}$ contains even number of Green functions, while $\sigma^{\rm sk}_{xy}$ consists of odd number of Green functions.
 Hence, the total Hall conductivity for the cases (i) and (ii) is no longer (anti)symmetric for the chemical potential (\cref{fig:AHE-wVC}).
 This feature is shared by the spin Hall effect in the non-magnetic Dirac electron system~\cite{Fukazawa2017}.

%
%
%

 As pointed out in Ref.~\cite{Fujimoto2014}, the finite Hall conductivities are obtained in the band gap for the cases of (ii) and (iii) because of $\sigma_{xy}^{\mathrm{sea}}$, and their values depend on the way of the momentum cut-off.
 We take isotropic cut-off in momentum space in \cref{fig:AHE-wVC}.
 In the paper by Goswami~\cite{Goswami2013}, the anomalous Hall conductivity in a topological phase and the similar dependences on the momentum cut-off are discussed.
 They conclude that the regularization by using the cylindrical momentum cut-off is reasonable on the basis of the bulk-boundary correspondence.
 As the system we consider is in the trivial phase and there is no such correspondence, it is not clear how to take the regularization.
 For more realistic situations, it may depend on a system which we consider.

\section{\label{sec:conclusion}Conclusion}
 In conclusion, we investigate the effects of the VCs on the conductivity tensor of the Dirac ferromagnet.
 By considering the VCs, the diagonal conductivities increase, and the increments are understood as the renormalization of the lifetime and the contributions from the correlations between the charge- and spin-currents.
 However, the AMR does not change quantitatively because the VCs contribute almost equally to the conductivities parallel and perpendicular to the ferromagnetic order parameters.
 For the AHE, the extrinsic contributions such as the side-jump and skew-scattering ones are calculated, and the skew-scattering contribution is dominant in the clean case, as seen in the spin Hall effect in the non-magnetic 3D Dirac electron system~\cite{Fukazawa2017}.

\begin{acknowledgments}
 The author would like to thank H.~Kohno and G.~Tatara for valuable discussion, and A.~Shitade for giving informative comments.
 This work was supported by a Grant-in-Aid for Specially Promoted Research (No.~15H05702).
\end{acknowledgments}

\onecolumngrid

\appendix
\section{\label{apx:correlation} Intermixing of particle-, `magnetization'- and `spin'-currents mediated by impurity scatterings}
 In this appendix, we calculate the first order VCs to the velocity vertexes of particle-, `magnetization'- and `spin'-current.
 Then, we show that the impurity scatterings cause an intermixing between the particle-current and the `magnetization'-current in the $\hat{x}, \hat{y}$-direction, and between the particle-current and `spin'-current in the $\hat{z}$-direction.

\subsection{\label{apx:jxy-jMxy} Intermixing of particle- and `magnetization'-currents}

 The velocity vertexes of particle-, `magnetization- and `spin'-currents are given by \cref{eq:def_charge_current_velocity,eq:def_magnetization_current_velocity,eq:def_spin_current_velocity}.
 As the contributions to the conductivities in the leading order of the impurity concentration are of our interest, it is enough to evaluate
\begin{align}
\Ni u^2 \sum_{\bk} G^{\R}_{\bk} (\epsilon) \rho_{\mu} \sigma^{\nu} G^{\A}_{\bk} (\epsilon) \Big|_{\epsilon = \mu}
	& = \bbracket{ \mu \nu; \mu' \nu' } \rho_{\mu'} \sigma^{\nu'}
\label{eq:def_GR_rho_sigma_GA}
\end{align}
up to $\Ord{\Ni}$ at the Fermi level.
 In the right hand side, we expanded by using the two kinds of Pauli matrices, and the coefficients are given as
\begin{align}
\bbracket{ \mu \nu; \mu' \nu' }
	& = \frac{1}{4} \Ni u^2 \sum_{\bk} \tr \left[
		G^{\R}_{\bk} (\epsilon) \rho_{\mu} \sigma^{\nu} G^{\A}_{\bk} (\epsilon) \rho_{\mu'} \sigma^{\nu'}
	\right] \Big|_{\epsilon = \mu}
\label{eq:def_bbracket}
.\end{align}
 We first consider the cases of $v_i$ and $v^{z}_{M,i}$ for $i = x$ and $i = y$.
 Substituting $\mu = 1, 2$ and $\nu = x, y$ into \cref{eq:def_bbracket}, and taking the traces, we find that almost all the coefficients of $\rho_{\mu'} \sigma^{\nu'}$ in \cref{eq:def_GR_rho_sigma_GA} vanish, and the non-vanishing components can be collectively expressed as
\begin{align}
\Ni u^2 \sum_{\bk} G^{\R}_{\bk} (\epsilon)
	\begin{pmatrix}
		\rho_1 \sigma^x
	\\	\rho_2 \sigma^y
	\\	\rho_1 \sigma^y
	\\	\rho_2 \sigma^x
	\end{pmatrix}
G^{\A}_{\bk} (\epsilon) \Big|_{\epsilon = \mu}
 & = 
	\begin{pmatrix}
		\bbracket{1x;1x}
	&	\bbracket{1x;2y}
	&	\bbracket{1x;1y}
	&	\bbracket{1x;2x}
	\\	\bbracket{2y;1x}
	&	\bbracket{2y;2y}
	&	\bbracket{2y;1y}
	&	\bbracket{2y;2x}
	\\	\bbracket{1y;1x}
	&	\bbracket{1y;2y}
	&	\bbracket{1y;1y}
	&	\bbracket{1y;2x}
	\\	\bbracket{2x;1x}
	&	\bbracket{2x;2y}
	&	\bbracket{2x;1y}
	&	\bbracket{2x;2x}
	\end{pmatrix}
	\begin{pmatrix}
		\rho_1 \sigma^x
	\\	\rho_2 \sigma^y
	\\	\rho_1 \sigma^y
	\\	\rho_2 \sigma^x
	\end{pmatrix}
\\
	& = \left( \hat{A}_{\perp}^{(0)} + \hat{A}_{\perp}^{(1)} \right)
	\begin{pmatrix}
		\rho_1 \sigma^x
	\\	\rho_2 \sigma^y
	\\	\rho_1 \sigma^y
	\\	\rho_2 \sigma^x
	\end{pmatrix}
\label{eq:GR_xy_GA}
,\end{align}
where $\hat{A}_{\perp}^{(0)}$ and $\hat{A}_{\perp}^{(1)}$ are $\Ord{\Ni^0}$- and $\Ord{\Ni}$-order terms, as we will show below.
 Here, $\bbracket{1x;1x}$ (and also $\bbracket{1y;1y}$) is proportional to the diagonal charge conductivity evaluated from the bare-bubble diagram (see \cref{eq:jx-jx}), $\bbracket{1x;2y}$ and $\bbracket{1y;2x}$ (also $\bbracket{2y;1x}$ and $\bbracket{2x;1y}$) are proportional to the correlation functions between the (diagonal) charge-current and the `magnetization'-current.
 In order to emphasize this point, we write $\hat{A}_{\perp}^{(0)}$ as
\begin{align}
\hat{A}_{\perp}^{(0)}
	& =
	\begin{pmatrix}
		\bbracket{j_x; j_x}
	&	- \bbracket{j_x; j^{z}_{M,x}}
	&	0
	&	0
	\\	- \bbracket{j^{z}_{M,x}; j_x}
	&	\bbracket{j^{z}_{M,x}; j^{z}_{M,x}}
	&	0
	&	0
	\\	0
	&	0
	&	\bbracket{j_y; j_y}
	&	\bbracket{j_y; j^{z}_{M,y}}
	\\	0
	&	0
	&	\bbracket{j^{z}_{M,y}; j_y}
	&	\bbracket{j^{z}_{M,y}; j^{z}_{M,y}}
	\end{pmatrix}
\label{eq:A0_perp}
,\end{align}
where these coefficients are obtained as
\begin{subequations}
\begin{align}
\bbracket{j_x; j_x}
	& = \bbracket{j_y; j_y}
	= \bbracket{1x; 1x}
	\simeq \overline{ k_{\perp}^2 \Delta_{\bm{k}}^2 }
	= \frac{\pi}{2} \frac{\Ni u^2}{e^2} \sigma^{\mathrm{b}}_{\perp}
\label{eq:jx-jx}
, \\
\bbracket{j^{z}_{M,x}; j^{z}_{M,x}}
	& = \bbracket{j^{z}_{M,y}; j^{z}_{M,y}}
	= \bbracket{2 x; 2 x}
	\simeq \Omega^2 \overline{ k_{\perp}^2 }
\label{eq:jMx-jMx}
, \\
\bbracket{j_x; j^{z}_{M,x}}
	& = \bbracket{j^{z}_{M,x}; j_x}
	= - \bbracket{1x; 2y}
	= - \bbracket{j_y; j^{z}_{M,y}}
	= - \bbracket{j^{z}_{M,y}; j_y}
	\simeq - \Omega \overline{ \eta k_{\perp}^2 \Delta_{\bm{k}} }
\label{eq:jx-jMx}
.\end{align}
\label{eq:A_perp^0-components}%
\end{subequations}
 We introduced the expression, $\overline{X (\bk, \eta)}$ of a certain function $X (\bk, \eta)$, as
\begin{align}
\overline{X (\bk,\eta)}
	& = \frac{\pi}{4} \Ni u^2
	\sum_{\bm{k}, \eta} \Theta_{\eta} (\mu)
	\frac{ 1 }{\Delta_{\bm{k}}}
	\frac{ X (\bm{k}, \eta) }{| \eta \Delta_{\bm{k}} \Gamma_1 - \Omega \Gamma_2 - k_z^2 \Gamma_3 |}
	\delta ( k_{\perp}^2 - \alpha_{\eta} )
,\end{align}
and we neglected the higher order in $\Ni$.
 Here, $\Gamma_j$ ($j = 1,2,3$), $\alpha_{\eta}$, and $\Theta_{\eta} (\mu)$ are defined respectively by Eqs.(20)-(22), Eq.~(41), and Eq.~(42) in Ref.~\cite{Fujimoto2014}.
 The $k_{\perp}$-integrals are performed by analytically, and the $k_z$-integrals are numerically caluclated.

 Similar relations can be found in $\hat{A}_{\perp}^{(1)}$, that $\bbracket{1x,1y}$ (and $\bbracket{1y,1x}$) is the Fermi-surface contribution to the off-diagonal charge conductivity evaluated from the bare-bubble diagram.
 However, the Fermi-sea contribution gives rise to an important contribution to off-diagonal charge conductivity $\bbracket{j_x; j_y}$, and $\bbracket{1x,1y}$ is not equivalent to $\bbracket{j_x; j_y}$.
 To keep this difference obvious, $\bbracket{1x,1y}$ is denote by $\bbracketsurf{j_x; j_y}$.
 Then, $\hat{A}_{\perp}^{(1)}$ is given as
\begin{align}
\hat{A}_{\perp}^{(1)}
	& =
	\begin{pmatrix}
		0
	&	0
	&	\bbracketsurf{j_x; j_y}
	&	\bbracketsurf{j_x; j^{z}_{M,y}}
	\\	0
	&	0
	&	- \bbracketsurf{j^{z}_{M,x}; j_y}
	&	- \bbracketsurf{j^{z}_{M,x}; j^{z}_{M,y}}
	\\	\bbracketsurf{j_y; j_x}
	&	- \bbracketsurf{j_y; j^{z}_{M,x}}
	&	0
	&	0
	\\	\bbracketsurf{j^{z}_{M,y}; j_x}
	&	- \bbracketsurf{j^{z}_{M,y}; j^{z}_{M,x}}
	&	0
	&	0
	\end{pmatrix}
\label{eq:A1_perp}
,\end{align}
where 
\begin{subequations}
\begin{align}
\bbracketsurf{j_x; j_y}
	& = - \bbracketsurf{j_y; j_x}
	= \bbracket{1x;1y}
	\simeq - 2 \overline{ \eta \Delta_{\bm{k}} C_{xy} }
\label{eq:jx-jy}
, \\
\bbracketsurf{j^{z}_{M,x}; j^{z}_{M,y}}
	& = - \bbracketsurf{j^{z}_{M,y}; j^{z}_{M,x}}
	= \bbracket{2x;2y}
	\simeq - 2 \overline{ ( \eta \Delta_{\bm{k}} + k_z^2 ) C_{xy}} - 2 (S \gamma_{0 0} - M \gamma_{3 0}) \overline{k_z^2 k_{\perp}^2 }
\label{eq:jMx-jMy}
, \\
\bbracketsurf{j_x; j^{z}_{M,y}}
	& = - \bbracketsurf{j^{z}_{M,y}; j_x}
	= \bbracketsurf{j^{z}_{M,x}; j_y}
	= - \bbracket{j_y; j^{z}_{M,x}}
	= \bbracket{1x;2x}
	\simeq 2 \Omega \overline{ C_{xy}} - ( M \gamma_{0 z} - S \gamma_{3 z}) \overline{k_z^2 k_{\perp}^2 }
\label{eq:jx-jMy}
,\end{align}
\label{eq:A_perp^1-components}
\end{subequations}
and $C_{xy}$ is defined by Eq.~(51) in Ref.~\cite{Fujimoto2014}.

\subsection{\label{apx:jz-jSz} Intermixing of particle- and `spin'-currents}
 Second, we calculate \cref{eq:def_GR_rho_sigma_GA} in the cases of $v_i$ and $v_{S,i}^{z}$ for $i = z$.
 Since AHE does not arise in this direction, it is sufficient to evaluate the VCs up to $\Ord{\Ni^0}$.
 Substituting $\mu = 1$, $\nu = 0, z$ into \cref{eq:def_bbracket}, and we find the similar expressions as $i = x, y$ as
\begin{align}
\Ni u^2 \sum_{\bk} G^{\R}_{\bk} (\epsilon)
	\begin{pmatrix}
		\rho_1 \sigma^z
	\\	\rho_1 \sigma^0
	\end{pmatrix}
G^{\A}_{\bk} (\epsilon) \Big|_{\epsilon = \mu}
=
	\hat{A}_{\parallel}^{(0)}
	\begin{pmatrix}
		\rho_1 \sigma^z
	\\	\rho_1 \sigma^0
	\end{pmatrix}
	+ \Ord{\Ni} (\rho_2 \sigma^0 + \rho_2 \sigma^z)
\label{eq:GR_1z_10_GA}
,\end{align}
where the coefficient matrix $\hat{A}_{\parallel}^{(0)}$ is $\Ord{\Ni^0}$, 
\begin{align}
\hat{A}_{\parallel}^{(0)}
	& =
	\begin{pmatrix}
		\bbracket{j_z; j_z}
	&	\bbracket{j_z; j^{z}_{S,z}}
	\\	\bbracket{j^{z}_{S,z}; j_z}
	&	\bbracket{j^{z}_{S,z}; j^{z}_{S,z}}
	\end{pmatrix}
\label{eq:A0_parallel}
,\end{align}
and the matrix elements up to $\Ord{\Ni^0}$ are given by
\begin{subequations}
\begin{align}
\bbracket{j_z; j_z}
	= \bbracket{1z; 1z}
	& \simeq 2 \overline{ k_z^2 ( \eta \Delta_{\bm{k}} + S^2 - M^2)^2 }
	= \frac{\pi}{2} \frac{\Ni u^2}{e^2} \sigma_{\parallel}^{\mathrm{b}}
\label{eq:jz-jz}
, \\
\bbracket{j^{z}_{S,z}; j^{z}_{S,z}}
	= \bbracket{10; 10}
	& \simeq 2 (\mu S + m M)^2 \overline{ k_z^2 }
\label{eq:jSz-jSz}
, \\
\bbracket{j_z; j^{z}_{S,z}}
	& = \bbracket{j^{z}_{S,z}; j_z}
	= \bbracket{1z; 10}
	\simeq - 2 (\mu S + m M) \overline{ \eta k_z^2 \Delta_{\bm{k}} }
\label{eq:jz-jSz}
.\end{align}
\label{eq:A0_parallel^0-components}%
\end{subequations}
 \Cref{eq:A0_parallel} shows that the particle-current and the `spin'-current in the $\hat{z}$-direction are mixed by the impurity scatterings.

\section{\label{apx:vc}Ladder type VCs for velocity vertexes of particle-, `magnetization'- and `spin'-current}
 Next, we calculate the ladder type of VCs.
 From \cref{eq:def_GR_rho_sigma_GA}, it is easy to extend to arbitrary order VCs.
 For example, the VC of the second order is calculated as
\begin{align}
(\Ni u^2)^2 \sum_{\bk, \bk'}
&	G^{\R}_{\bk'} (\epsilon) G^{\R}_{\bk} (\epsilon)
		\rho_{\mu} \sigma^{\nu}
	G^{\A}_{\bk} (\epsilon) G^{\A}_{\bk'} (\epsilon) \Big|_{\epsilon = \mu}
\notag \\
	& = \bbracket{ \mu \nu; \mu' \nu' }
	\Ni u^2 \sum_{\bk'} G^{\R}_{\bk'} (\epsilon) \rho_{\mu'} \sigma^{\nu'} G^{\A}_{\bk'} (\epsilon) \Big|_{\epsilon = \mu}
\notag \\
	& = \bbracket{ \mu \nu; \mu' \nu' } \bbracket{ \mu' \nu'; \mu'' \nu''} \rho_{\mu''} \sigma^{\nu''}
\label{eq:second_VC}
.\end{align}
 By using \cref{eq:GR_xy_GA,eq:second_VC}, we can calculate the ladder type VCs to the velocities of the particle- and `magnetization'-current in the $x$- and $y$-directions as
\begin{align}
	\begin{pmatrix}
		\tilde{\varLambda}_{1,x}
	\\	\tilde{\varLambda}_{2,y}
	\\	\tilde{\varLambda}_{1,y}
	\\	\tilde{\varLambda}_{2,x}
	\end{pmatrix}
	& =
	\left[
	\hat{1} + \hat{A}_{\perp}^{(0)} + \hat{A}_{\perp}^{(1)} + \bigl(\hat{A}_{\perp}^{(0)} + \hat{A}_{\perp}^{(1)} \bigr)^2 + \cdots
	\right]
	\begin{pmatrix}
		\rho_1 \sigma^x
	\\	\rho_2 \sigma^y
	\\	\rho_1 \sigma^y
	\\	\rho_2 \sigma^x
	\end{pmatrix}
\notag \\
	& \simeq
	\left[
	\sum_{n = 0}^{\infty} \left(\hat{A}_{\perp}^{(0)}\right)^n
	+ \sum_{n,m = 0}^{\infty} \left(\hat{A}_{\perp}^{(0)}\right)^n \hat{A}_{\perp}^{(1)} \left(\hat{A}_{\perp}^{(0)}\right)^m
	\right]
	\begin{pmatrix}
		\rho_1 \sigma^x
	\\	\rho_2 \sigma^y
	\\	\rho_1 \sigma^y
	\\	\rho_2 \sigma^x
	\end{pmatrix}
\label{eq:vc_xy_calculated}
,\end{align}
where $\hat{A}_{\perp}^{(m)}$ ($m = 0, 1$) is $\Ord{\Ni^m}$-term, and we dropped $\Ord{\Ni^2}$.
 The first term is further calculated as
\begin{align}
\sum_{n = 0}^{\infty} \left(\hat{A}_{\perp}^{(0)}\right)^n
	& = 
	\frac{1}{\mathcal{D}_{\perp} }
	\begin{pmatrix}
		1 - \bbracket{j^{z}_{M,x}; j^{z}_{M,x}}
	&	- \bbracket{j_x; j^{z}_{M,x}}
	&	0
	&	0
	\\	- \bbracket{j_x; j^{z}_{M,x}}
	&	1 - \bbracket{j_x; j_x}
	&	0
	&	0
	\\	0
	&	0
	&	1 -\bbracket{j^{z}_{M,x}; j^{z}_{M,x}}
	&	\bbracket{j_x; j^{z}_{M,x}}
	\\	0
	&	0
	&	\bbracket{j_x; j^{z}_{M,x}}
	&	1 - \bbracket{j_x; j_x}
	\end{pmatrix}
\label{eq:tilde_Lambda_xy_Order_0}
,\end{align}
where
\begin{align}
\mathcal{D}_{\perp}
	& = \left( 1 - \bbracket{j_x; j_x} \right) \left( 1 - \bbracket{j^{z}_{M,x}; j^{z}_{M,x}} \right)
		 - \bbracket{j_x; j^{z}_{M,x}}^2
\tag{\ref{eq:def_D_perp}}
.\end{align}
 We also obtain the second term of \cref{eq:vc_xy_calculated} as
\begin{align}
\sum_{n,m = 0}^{\infty} \left(\hat{A}_{\perp}^{(0)}\right)^n \hat{A}_{\perp}^{(1)} \left(\hat{A}_{\perp}^{(0)}\right)^m
	& =
	\frac{1}{ ( \mathcal{D}_{\perp} )^2 }
	\begin{pmatrix}
		0
	&	0
	&	- \mathcal{P}_{pp}
	&	- \mathcal{P}_{Mp}
	\\	0
	&	0
	&	\mathcal{P}_{Mp}
	&	\mathcal{P}_{MM}
	\\	\mathcal{P}_{pp}
	&	- \mathcal{P}_{Mc}
	&	0
	&	0
	\\	\mathcal{P}_{Mp}
	&	- \mathcal{P}_{MM}
	&	0
	&	0
	\end{pmatrix}
\label{eq:tilde_Lambda_xy_Order_1}
\end{align}
with the matrix elements given by
\begin{align}
\mathcal{P}_{pp}
	& = 2 \bbracket{j_x; j^{z}_{M,x}} \left( 1 - \bbracket{j^{z}_{M,x}; j^{z}_{M,x}} \right) \bbracketsurf{j_x; j^{z}_{M,y}}
		- \left( 1 - \bbracket{j^{z}_{M,x}; j^{z}_{M,x}} \right)^2 \bbracketsurf{j_x; j_y}
\notag \\ & \hspace{2em}
		- \bbracket{j_x; j^{z}_{M,x}}^2 \bbracketsurf{j^{z}_{M,x}; j^{z}_{M,y}}
\label{eq:P_pp}
, \\
\mathcal{P}_{Mp}
	& = \bbracket{j_x; j^{z}_{M,x}} \left\{ \bbracketsurf{j_x; j_y} (1-\bbracket{j^{z}_{M,x}; j^{z}_{M,x}}) + \bbracketsurf{j^{z}_{M,x}; j^{z}_{M,y}} (1-\bbracket{j_x; j_x}) \right\}
\notag \\ & \hspace{2em}
		- \bbracketsurf{j_x; j^{z}_{M,y}} \left\{ \left( 1 - \bbracket{j_x; j_x} \right) \left( 1 - \bbracket{j^{z}_{M,x}; j^{z}_{M,x}} \right) + \bbracket{j_x; j^{z}_{M,x}}^2 \right\}
, \\
\mathcal{P}_{MM}
	& = 2 (1-\bbracket{j_x; j_x}) \bbracket{j_x; j^{z}_{M,x}} \bbracketsurf{j_x; j^{z}_{M,y}} - \bbracket{j_x; j^{z}_{M,x}}^2 \bbracketsurf{j_x; j_y} - (1-\bbracket{j_x; j_x})^2 \bbracketsurf{j^{z}_{M,x}; j^{z}_{M,y}}
.\end{align}

 The ladder type VCs to the velocities of charge- and `spin'-current in the $\hat{z}$-direction are also obtained as
\begin{align}
	\begin{pmatrix}
		\tilde{\varLambda}_{1,z}
	\\	\tilde{\varLambda}_{1,0}
	\end{pmatrix}
	& \simeq \sum_{n=0}^{\infty} \bigl( \hat{A}_{\parallel}^{(0)} \bigr)^n
	\begin{pmatrix}
		\rho_1 \sigma^z
	\\	\rho_1 \sigma^0
	\end{pmatrix}
	= \frac{1}{\mathcal{D}_{\parallel}}
	\begin{pmatrix}
		1 - \bbracket{j^{z}_{S,z}; j^{z}_{S,z}}
	&	\bbracket{j_z; j^{z}_{S,z}}
	\\	\bbracket{j_z; j^{z}_{S,z}}
	&	1 - \bbracket{j_z; j_z}
	\end{pmatrix}
	\begin{pmatrix}
		\rho_1 \sigma^z
	\\	\rho_1 \sigma^0
	\end{pmatrix}
\label{eq:vc_z0_calculated}
,\end{align}
where
\begin{align}
\mathcal{D}_{\parallel}
	& = \left( 1 - \bbracket{j_z; j_z} \right) \left( 1 - \bbracket{j^{z}_{S,z}; j^{z}_{S,z}} \right)
		- \bbracket{j_z; j^{z}_{S,z}}^2
\tag{\ref{eq:def_D_para}}
.\end{align}

\section{\label{apx:conductivity}Calculation of the conductivity tensor}
 As we have obtained $\tilde{\varLambda}_{1,i}$ ($i = x, y, z$) in \cref{apx:vc}, we here perform the calculation of the conductivity tensor.
\subsection{\label{apx:longitudinal}Diagonal conductivity}
 First, the diagonal conductivities [\cref{eq:AMR-wVC}] are calculated as
\begin{align}
\sigma_{i i}^{}
	& =
		\frac{e^2}{2 \pi \Ni u^2} 
		\tr \left[
			\rho_1 \sigma^i
			\left( \Ni u^2\sum_{\bk} G_{\bk}^{\R} (\epsilon) \tilde{\varLambda}_{1,i} G_{\bm{k}}^{\A} (\epsilon) \bigl. \Bigr|_{\epsilon=\mu} \right)
		\right]
\notag \\
	& = 
		\frac{e^2}{2 \pi \Ni u^2} 
		\tr \left[
			\rho_1 \sigma^i
			\left( \tilde{\varLambda}_{1,i} - \rho_1 \sigma^i \right)
		\right]
,\end{align}
where we have used \cref{eq:varLambda_sj}.

 For $i = x$, substituting \cref{eq:vc_xy_calculated} with \cref{eq:tilde_Lambda_xy_Order_0} and taking the trace, we obtain
\begin{align}
\sigma_{\perp}
	& = \sigma_{xx}^{}
	= \frac{2 e^2}{\pi \Ni u^2} \frac{1}{\mathcal{D}_{\perp}}
	\left\{
		\Bigl( 1 - \bbracket{j^{z}_{M,x}; j^{z}_{M,x}} \Bigr) \bbracket{j_{x}; j_{x}}
		+ \bbracket{j_x; j^{z}_{M,x}}^2
	\right\} 
,\end{align}
where $\mathcal{D}_{\perp}$ is given by \cref{eq:def_D_perp}, and $\bbracket{j^{}_{x}; j^{}_{x}}$, $\bbracket{j^{}_{x}; j^{z}_{M,x}}$ and $\bbracket{j^{z}_{M,x}; j^{z}_{M,x}}$ are evaluated as \cref{eq:A_perp^0-components}.
 By using \cref{eq:result_sigma_perp_woVC} and $\bbracket{j^{}_{x}; j^{z}_{M,x}} = \bbracket{j^{z}_{M, x}; j^{}_{x}}$, we obtain the result for $\sigma_{\perp}$ as shown in \cref{eq:result_sigma} with \cref{eq:result_sigma_perp,eq:result_sigma_add_perp}.
 Similar procedures are performed for $i = y$, and it is found $\sigma_{yy} = \sigma_{xx}$ as we expected from the symmetry.

 For $i = z$, substituting \cref{eq:vc_z0_calculated} and taking the trace,
\begin{align}
\sigma_{\parallel}
	& = \sigma_{zz}
	= \frac{2 e^2}{\pi \Ni u^2} \frac{1}{\mathcal{D}_{\parallel}}
	\left\{
		\Bigl( 1 - \bbracket{j^{z}_{S,z}; j^{z}_{S,z}} \Bigr) \bbracket{j_{z}; j_{z}}
		+ \bbracket{j_z; j^{z}_{S,z}}^2
	\right\}
\end{align}
is obtained as shown in \cref{eq:result_sigma} with \cref{eq:result_sigma_para,eq:result_sigma_add_para}.
 Here, $\mathcal{D}_{\parallel}$ is given by \cref{eq:def_D_para}, and $\bbracket{j_{z}; j_{z}}$, $\bbracket{j_z; j^{z}_{S,z}}$ and $\bbracket{j^{z}_{S,z}; j^{z}_{S,z}}$ are shown in \cref{eq:A0_parallel^0-components}.

\subsection{\label{apx:transverse_ladder}Contribution from ladder type VCs to off-diagonal conductivity}
 The off-diagonal conductivity including the bare-bubble contribution and the ladder type VCs [\cref{eq:AHE-b+sj}] is calculated as similar to the procedure of the diagonal one, 
\begin{align}
\sigma_{xy}^{\rm b+sj}
	& =
		\frac{e^2}{2 \pi \Ni u^2} 
		\tr \left[
			\rho_1 \sigma^x
			\left( \Ni u^2\sum_{\bk} G_{\bk}^{\R} (\epsilon) \tilde{\varLambda}_{1,y} G_{\bm{k}}^{\A} (\epsilon) \bigl. \Bigr|_{\epsilon=\mu} \right)
		\right]
\notag \\
	& = 
		\frac{e^2}{2 \pi \Ni u^2} 
		\tr \left[
			\rho_1 \sigma^x
			\tilde{\varLambda}_{1,y}
		\right]
,\end{align}
where we used \cref{eq:varLambda_sj} and $\tr [ \rho_1 \sigma^x \rho_1 \sigma^y] = 0$.
 Substituting \cref{eq:vc_xy_calculated,eq:tilde_Lambda_xy_Order_0,eq:tilde_Lambda_xy_Order_1}, all the components of \cref{eq:tilde_Lambda_xy_Order_0} vanish because of the trace, and the $\rho_1 \sigma^x$-component of \cref{eq:tilde_Lambda_xy_Order_1} only remains,
\begin{align}
\sigma_{xy}^{\rm b+sj}
	& = \frac{2 e^2}{\pi \Ni u^2}
		\frac{ \mathcal{P}_{pp} }{ ( \mathcal{D}_{\perp} )^2 }
,\end{align}
where $\mathcal{P}_{pp}$ is given by \cref{eq:P_pp}.
 By using \cref{eq:result_AHE_woVC,eq:A_perp^0-components}, we obtain \cref{eq:result_AHE_b+sj} with \cref{eq:result_AHE_b_tilde,eq:result_AHE_additional}.

\subsection{\label{apx:transverse_skew}Contribution from skew-scattering type VCs to off-diagonal conductivity}
 Finally, we calculate the contribution form the skew-scattering type VCs [\cref{eq:AHE-sk}].
 We here evaluate $\tilde{\varLambda}^{*}_{1,i}$ ($i = x,y$) defined by interchanging $\R$ and $\A$ in $\tilde{\varLambda}_{1,i}$.
 Using \cref{eq:def_GR_rho_sigma_GA} and 
\begin{align}
\Ni u^2 \sum_{\bk} G^{\A}_{\bk} (\epsilon) \rho_{\mu} \sigma^{\nu} G^{\R}_{\bk} (\epsilon) \Big|_{\epsilon = \mu}
	& = \bbracket{ \mu' \nu'; \mu \nu } \rho_{\mu'} \sigma^{\nu'}
,\end{align}
we find that the $\Ord{\Ni^0}$-terms of $\tilde{\varLambda}^{*}_{1,i}$ are equivalent to those of $\tilde{\varLambda}^{}_{1,i}$, hence, $\tilde{\varLambda}^{*}_{1,i} = \tilde{\varLambda}_{1,i}$ in the leading order.
 We calculate \cref{eq:AHE-sk} as
\begin{align}
\sigma_{x y}^{\mathrm{sk}}
	& = \frac{e^2}{4 \pi \Ni^2 u^3}
		\mathrm{tr} \left[
			\left( \Ni u^2 \sum_{\bk}
			 G^{\A}_{\bk}
			\tilde{\varLambda}^{*}_{1,x}
			 G^{\R}_{\bk}
			 \right)
			\left( \Ni u^2 \sum_{\bk''}
			 G^{\R}_{\bk''}
			 \right)
			\left( \Ni u^2 \sum_{\bk'}
			 G^{\R}_{\bk'}
			\tilde{\varLambda}_{1,y}
			 G^{\A}_{\bk'}
			 \right)
\right. \notag \\ & \hspace{3em} \left.
			+ \left( \Ni u^2 \sum_{\bk}
			 G^{\A}_{\bk}
			\tilde{\varLambda}^{*}_{1,x}
			 G^{\R}_{\bk}
			 \right)
			\left( \Ni u^2 \sum_{\bk'}
			 G^{\R}_{\bk'}
			\tilde{\varLambda}_{1,y}
			 G^{\A}_{\bk'}
			 \right)
			\left( \Ni u^2 \sum_{\bk''}
			 G^{\A}_{\bk''}
			 \right)
		- ( \R \leftrightarrow \A )
		\right]
		\Bigr|_{\epsilon = \mu}
\notag \\
	& \simeq \frac{e^2}{2 \pi \Ni^2 u^3}
		\mathrm{tr} \left[
			\left( \tilde{\varLambda}_{1,x} - \rho_1 \sigma^x \right)
			\left( - \zi \gamma_{\mu \nu} \rho_{\mu} \sigma^{\nu} \right)
			\left( \tilde{\varLambda}_{1,y} - \rho_1 \sigma^y \right)
			+ \left( \tilde{\varLambda}_{1,x} - \rho_1 \sigma^x \right)
			\left( \tilde{\varLambda}_{1,y} - \rho_1 \sigma^y \right)
			\left( \zi \gamma_{\mu \nu} \rho_{\mu} \sigma^{\nu} \right)
		\right]
		\Bigr|_{\epsilon = \mu}
\notag \\
	& = \frac{e^2 \zi \gamma_{\mu \nu}}{2 \pi \Ni^2 u^3}
		\mathrm{tr} \left[
			\left\{
			\left( \tilde{\varLambda}_{1,x} - \rho_1 \sigma^x \right)
			\left( \tilde{\varLambda}_{1,y} - \rho_1 \sigma^y \right)
			- \left( \tilde{\varLambda}_{1,y} - \rho_1 \sigma^y \right)
			\left( \tilde{\varLambda}_{1,x} - \rho_1 \sigma^x \right)
			\right\}
			\rho_{\mu} \sigma^{\nu}
		\right]
		\Bigr|_{\epsilon = \mu}
.\end{align}
 Here,
\begin{align}
\{ \,\cdots \}
	& =
	\left( \bbracketld{j_x; j_x}\, \rho_1 \sigma^x - \bbracketld{j_x; j^{z}_{M,x}}\, \rho_2 \sigma^y \right)
	\left( \bbracketld{j_x; j_x}\, \rho_1 \sigma^y + \bbracketld{j_x; j^{z}_{M,x}}\, \rho_2 \sigma^x \right)
\notag \\ &
	- 
	\left( \bbracketld{j_x; j_x}\, \rho_1 \sigma^y + \bbracketld{j_x; j^{z}_{M,x}}\, \rho_2 \sigma^x \right)
	\left( \bbracketld{j_x; j_x}\, \rho_1 \sigma^x - \bbracketld{j_x; j^{z}_{M,x}}\, \rho_2 \sigma^y \right)
\notag \\
	& = 2 \zi \left(
		\bbracketld{j_x; j_x}^2 \sigma^z
		+ 2 \bbracketld{j_x; j_x} \bbracketld{j_x; j^{z}_{M,x}} \rho_3
		+ \bbracketld{j_x; j^{z}_{M,x}}^2 \sigma^z
		\right)
,\end{align}
and then we obtain \cref{eq:result_AHE_skew}.


\bibliography{references}
\end{document}